\documentclass[twocolumn,showpacs,preprintnumbers,amsmath,amssymb,prl]{revtex4}

\usepackage{graphicx}
\usepackage{dcolumn}
\usepackage{bm}
\DeclareMathVersion{chem}
\SetSymbolFont{letters}{chem}{OT1}{cmr}{m}{n}

\begin{document}


\title{High-Pressure Suppression of Long Range Magnetic Order\\
in the Triangular Lattice Antiferromagnet CuFeO$_{2}$}

\author{N. Terada$^{1,2}$
\email{TERADA.Noriki@nims.go.jp}}
\author{T. Osakabe$^{2}$}
\author{H. Kitazawa$^{1}$}
\affiliation{$^1$National Institute for Materials Science Sengen 1-2-1 Tsukuba Ibaraki 305-0047 Japan\\
$^2$Japan Atomic Research Agency Tokai Ibaraki 319-1195 Japan\\
}

\date{\today}

\begin{abstract}
We succeeded in observing pressure-suppressed magnetic long range ordering (LRO) in the triangular lattice antiferromagnet CuFeO$_{2}$, using neutron diffraction experiments under an isotropic pressure.
The magnetic LRO of the four-sublattice ground state under ambient pressure in CuFeO$_{2}$ almost disappears at the high pressure of 7.9 GPa, and is replaced by an incommensurate order with temperature-independent wave number of (0.192 0.192 1.5).
The incommensurate wave number observed at 7.9 GPa corresponds to that observed just above the temperature at which lattice distortion and magnetic LRO simultaneously occur under ambient pressure.
Therefore, the long-range magnetic ordering disappears because the high pressure suppressed the lattice distortion that otherwise relieves spin frustration and leads the spin system to LRO.

\end{abstract}

\maketitle
Spin-lattice coupling in frustrated magnetic materials is an important concept in understanding many unsolved cross-correlated phenomena such as multiferroics\cite{Kimura} and spontaneous distortions to release frustration.\cite{Matsuda,Yamashita,Tchernyshyov}
In frustrated magnets, lattice distortion often occurs to spontaneously remove the magnetic frustration.
Spontaneous lattice distortion is thus one of the essential factors for the realization of magnetic long range order (LRO) in frustrated magnetic systems.
Therefore, if the lattice distortion were suppressed by pressure, the magnetic LRO could not be stabilized and novel magnetic states such as spin liquids\cite{Gardner} and spin nematics\cite{Chandra,Nakatsuji} would be expected.
In the present study, we examined the effect of pressure on magnetic ordering in the frustrated spin-lattice coupling system of CuFeO$_{2}$.

Since the discovery of the spontaneous spin-lattice coupling phenomenon in the frustrated triangular lattice antiferromagnet CuFeO$_{2}$,\cite{N.Terada(2006-2),Ye} spin-lattice coupling has been extensively studied using high-field X-ray diffraction,\cite{N.Terada(2006-3),N.Terada(2007-2)} ultrasonic velocity,\cite{Quirion} magnetization measurements,\cite{Lummen} and Landau theory approaches.\cite{Plumer}
These studies have demonstrated that spontaneous spin-lattice coupling plays an essential role in the stabilization of the magnetic ground state of CuFeO$_{2}$.
Although additional high-pressure studies would assist in the further understanding of spin-lattice coupling in frustrated magnetic systems, and novel spin states were expected under high pressure, only a few high-pressure investigations of CuFeO$_{2}$ were performed.
Xu {\it et al.} studied the high pressure effect on the magnetic ordering of CuFeO$_{2}$ using $^{57}$Fe M\"ossbauer spectroscopy at pressures of up to 27 GPa and low temperatures.\cite{Xu,Xu2}
They observed a change in the internal fields above 6 GPa.
Takahashi {\it et al.} also investigated changes in the phase transition temperatures at up to 0.7 GPa, using magnetic susceptibility measurements.\cite{Takahashi}
They observed a slight shift in the phase transition temperature, but
the microscopic magnetic ordering and correlation have not yet been clarified.

CuFeO$_{2}$ belongs to the $R\bar{3}m$ space group at room temperature, where the lattice constants in hexagonal notation are $a=b=3.030\ {\rm \AA}$ and $c=17.144\ {\rm \AA}$ at ambient pressure, and $a=b=2.970\ {\rm \AA}$ and $c=17.056\ {\rm \AA}$ at 7.66 GPa.\cite{Zhao2}
CuFeO$_{2}$ has magnetic moments of the orbital singlet Fe$^{3+}$ ($S=5/2, L=0$) and nonmagnetic Cu$^{1+}$ and O$^{2-}$.
With decreasing temperature from the paramagnetic phase, the system enters a partially disordered (PD) phase at $T_{\rm N1}=$ 14 K, in which the collinear magnetic moments along the $c$ axis are sinusoidally modulated in space.
In the PD phase, the wave vector assigned as ($q$ $q$ $\frac{3}{2}$) depends on temperature for $0.195\le q \le 0.217$.\cite{S.Mitsuda(1998)}
In the phase transition at $T_{\rm N1}$, a spontaneous lattice distortion from rhombohedral ($R\bar{3}m$) to monoclinic ($C2/m$) occurs.\cite{Ye, N.Terada(2006-3)}
The crystal lattice deforms from the ``equilateral" triangle to ``isosceles" triangle, which separates the nearest neighbor exchange interactions into two different exchange interactions in the PD phase.\cite{N.Terada(2006-3)}
With decreasing temperature from the PD phase, another phase transition occurs at $T_{\rm N2}=$ 11 K into a four-sublattice (4SL) phase in which the collinear magnetic moments along the $c$ axis are ordered and the sequence is $\uparrow\uparrow\downarrow\downarrow$ in the $ab$ plane.\cite{S.Mitsuda(1998),Ye}
In the 4SL phase, the crystal lattice deforms from the ``isosceles" triangle to ``scalene" triangle, which separates the nearest neighbor exchange bonds into three different bonds.\cite{N.Terada(2006-2),N.Terada(2006-3)}
The scalene triangular lattice distortion lowers the exchange energy when the 4SL magnetic LRO is stabilized.
Since these lattice distortions are considerably anisotropic ones that lower the symmetry and release the spin frustration, isotropic hydrostatic pressure should suppress the anisotropic lattice distortions.

Recently, Osakabe {\it et al.} developed a {\it hybrid-anvil-type} high-pressure device for neutron scattering experiments.\cite{Osakabe}
The device has made possible the investigation of high-pressure effects on magnetic ordering with lattice distortion in frustrated magnetic systems.
In this study, we investigated the effects of high pressure on magnetic ordering in CuFeO$_{2}$ using neutron diffraction measurements at the high pressure of 7.9 GPa.
We found magnetic ordering with an incommensurate propagation wave vector (0.192 0.192 1.5) at the low temperature of 3 K and the high pressure of 7.9 GPa, instead of long range magnetic ordering with a commensurate wave vector of (0.25 0.25 1.5) at ambient pressure.

A single crystal of CuFeO$_{2}$ was grown by the floating zone method.\cite{T.R.Zhao(1996)}
For high-pressure measurement, the sample was cut to $0.6\times0.5\times0.25$ mm$^3$, and had a mass of 7.7 mg.
The hybrid-anvil device was described in detail in a previous paper.\cite{Osakabe}
We used glycerin as the pressure-transmitting medium because it transmits hydrostatic pressure well at higher pressures than other liquid media commonly used in high-pressure measurements.\cite{Osakabe2}
The pressure was determined at room temperature by ruby fluorescence.
We confirmed that the lattice constants at room temperature changed to the values reported in previous powder x-ray diffraction measurements under pressure.\cite{Zhao2}
The neutron diffraction measurements under high pressure and ambient pressure on CuFeO$_{2}$ were carried out with triple axis spectrometers TAS1 and TAS2 in JRR-3 of JAEA in Tokai, Japan.
The incident neutron energy was 30.5 meV, and no collimation was used, in order to maximize the neutron flux at the sample position.
In order to eliminate higher order contamination of the incident neutrons, we used a pyrolytic graphite filter. 
The sample was mounted in the hybrid-anvil device [1$\bar{1}0$] axis vertically, to provide access to the ($H H L$) scattering plane.
The high-pressure device was set to  closed-cycle $^4$He gas refrigerator.
We observed that the crystal mosaic gets worse under high pressure owing to slight nonhydrostaticity. 
We also performed the magnetic susceptibility measurements on the identical sample to one used in the high pressure neutron scattering measurements after the pressure measurements, in order to check the restoration of the magnetic properties.
The susceptibility measurements were performed with a commercial based SQUID magnetometer of Quantum Design.
 
\begin{figure}[t]
\begin{center}
\includegraphics[scale=0.5]{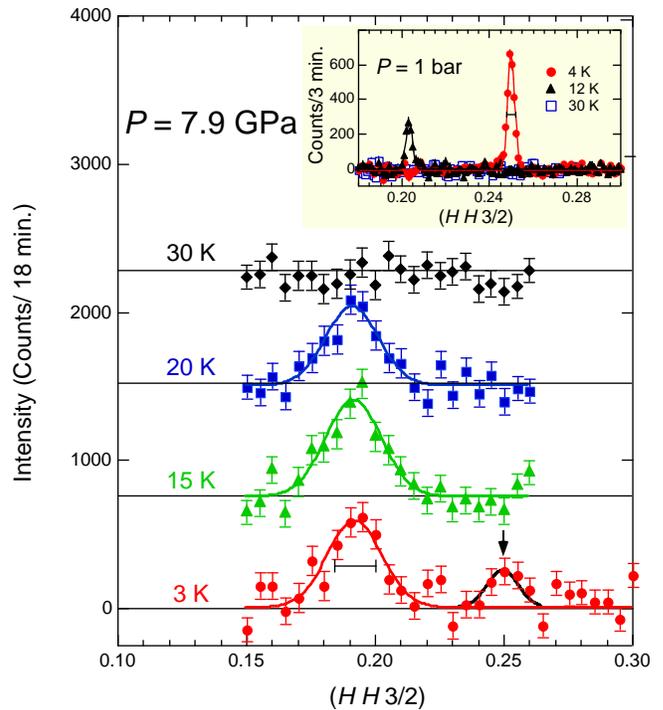}
\caption{(Color online) Neutron diffraction profiles of reciprocal lattice ($H H \frac{3}{2}$) scans at typical temperatures and at a pressure of 7.9 GPa in CuFeO$_{2}$.
The data at each temperature were subtracted by the data at 40 K.  
The solid lines denote the results of a Gaussian least squares fit.
The experimental resolution are represented by the horizontal bars. 
The arrow indicates the position of the commensurate position of (0.25 0.25 1.5)
The inset shows the data measured at an ambient pressure.}
\label{diffraction_profiles}
\end{center}
\end{figure}

The temperature dependence of the diffraction profiles at the high-pressure of 7.9 GPa is shown in Fig. \ref{diffraction_profiles}.
With decreasing temperature from the paramagnetic phase, magnetic scattering was observed at 20 K, 15 K, and 3 K.
These neutron scattering profiles are slightly broader than the experimental resolution, indicating that the magnetic correlation length is finite.
The peak position of this order is at incommensurate (0.192 0.192 1.5).
These results are unlike those under ambient pressure, where magnetic long range ordering was observed below 14 K and the incommensurate to commensurate transition occurred at 11 K, as shown in the inset of Fig. \ref{diffraction_profiles}.
As shown at the bottom of Fig. \ref{diffraction_profiles}, the magnetic signal was also observed at the commensurate position, which corresponds to the 4SL ground state at ambient pressure.
The integrated intensity of the commensurate components was four times smaller than that of the incommensurate components at (0.192 0.192 1.5).
Although the observed coexistence is not yet fully understood, we can conclude that the dominant component of the ground state did change, from commensurate LRO to incommensurate short range ordering (SRO).

\begin{figure}[htbp]
\begin{center}
\includegraphics[scale=0.55]{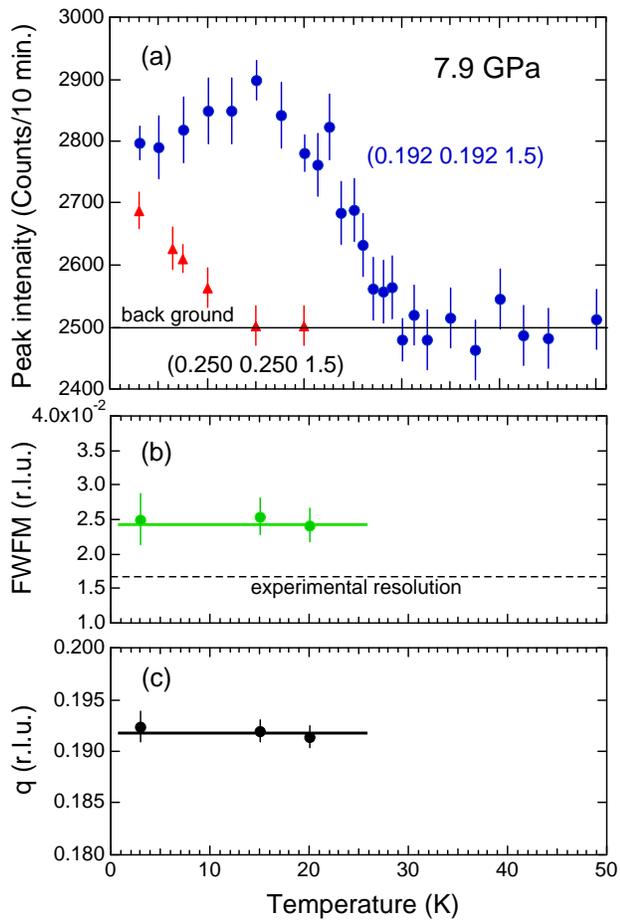}
\caption{(Color online) (a) Temperature dependence of the neutron peak intensity at the reciprocal lattice points (0.192 0.192 1.5) and (0.25 0.25 1.5) at 7.9 GPa.
The horizontal lines show the background.
Temperature dependence of (b) the full width at the half maximum, and (c) the wave number $q$ of the magnetic peak at ($q\ q\ \frac{3}{2}$).
The horizontal solid lines are visual guides, and the dotted line show the experimental resolution. 
}
\label{Temp_dep_int}
\end{center}
\end{figure}

As shown in Fig. \ref{Temp_dep_int}(a), magnetic scattering corresponding to the SRO appears below 30 K at 7.9 GPa.
This temperature is consistent with the previous M\"ossbauer measurements.\cite{Xu2}
Here, we call this temperature $T_{\rm HP}$ in the present paper, in order to distinguish it from $T_{\rm N1}$ and $T_{\rm N2}$ in ambient pressure.
With decreasing temperature from $T_{\rm HP}=$ 30 K, the intensity increased.
Below approximately 10 K, the intensity decreased slightly, and a magnetic peak at the commensurate position (0.25 0.25 1.5) appeared.
This temperature approximately corresponds to that of the appearance of the commensurate peak under ambient pressure.
The line width of the incommensurate peak is slightly broader than the experimental resolution and was independent of temperature, as shown in Fig. \ref{Temp_dep_int}(b).
The incommensurate wave number ($q\ q\ \frac{3}{2}$) is also independent of temperature, and $q=0.192$.

To investigate the restoration of magnetic LRO after the high-pressure neutron measurements, we performed the magnetic susceptibility measurements on the identical sample after the neutron measurements.
The observed susceptibility data are coincident with the data before the pressure measurements. (not shown)
We thus confirmed that the LRO is restored when the pressure is released.

\begin{figure}[t]
\begin{center}
\includegraphics[scale=0.55]{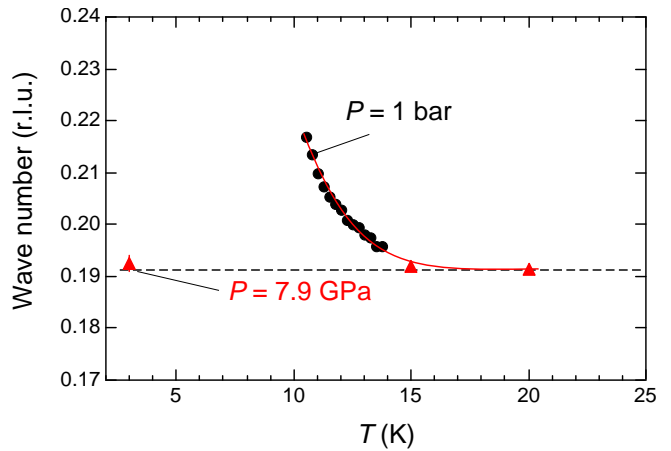}
\caption{(Color online) A comparison of the temperature dependence of the magnetic wave number between $P=7.9$ GPa and ambient pressure ($P=1$ bar). 
The circular and triangular symbols denote data at $P=7.9$ GPa and $P=1$ bar, respectively.
The solid curve and the horizontal line are visual guides.
The data at $P=1$ bar were taken from the previous paper.\cite{S.Mitsuda(1998)}}
\label{Tdep_q}
\end{center}
\end{figure}

We discuss the relationship between short range magnetic ordering at 7.9 GPa and long range ordering at ambient pressure with lattice distortion.
In previous synchrotron x-ray diffraction studies,\cite{Ye,N.Terada(2006-3)} crystal distortion occurred below $T_{\rm N1}=14$ K under ambient pressure.
Magnetic ordering with a long correlation length also occurred at $T_{\rm N1}$, and the spin system enters the PD phase.
In the PD phase, the magnetic wave vector depends upon temperature, as shown in Fig. \ref{Tdep_q}.
On the other hand, at $P=7.9$ GPa, the wave vector $(q\ q\ 1.5)$, with $q=0.192$, is independent of temperature.
The wave number at 7.9 GPa is almost the same as that of the SRO just above $T_{\rm N1}=14$ K under ambient pressure.
When the temperature dependence of $q$ under ambient pressure is extended to high temperatures above $T_{\rm N1}=14$ K, that merges the value around 0.192,  as shown by the solid curve in Fig. \ref{Tdep_q}.
Therefore, the high pressure strongly suppresses the PD state with temperature dependent wave number as well as the LRO in the 4SL state.
As discussed in previous papers,\cite{N.Terada(2006-2),Ye} the lattice distortion was believed to relieve the magnetic frustration, stabilizing the magnetic LRO.
We thus conclude that the magnetic long range ordering was prevented when the spontaneous lattice distortion was suppressed by the high pressure. 
To better understand this point, high-resolution neutron diffraction or x-ray diffraction measurements at high pressure would be helpful.

Xu {\it et al.} studied the effects of pressure on the magnetic ordering of CuFeO$_{2}$, using M\"ossbauer spectroscopy measurements up to 27 GPa.\cite{Xu,Xu2}
The M\"ossbauer data at 6 GPa indicated the coexistence of the 4SL state and another component.\cite{Xu}
At 19 GPa, they observed a spectrum that could only be explained by the presence of a single  static magnetic hyperfine field.
In the present neutron diffraction measurements, we observed SRO with an incommensurate wave vector at 7.9 GPa, which coexisted with the 4SL LRO.
Considering the results of the static hyperfine field in the M\"ossbauer study\cite{Xu} and the present neutron diffraction measurements, we find that the magnetic ordering observed under high pressure was a static ordering characterized by a short range magnetic correlation and an incommensurate wave vector.
Therefore, the magnetic entropy at high pressure was not fully released at the lowest temperature.
Inelastic neutron scattering and magnetic specific heat measurements under pressure for CuFeO$_{2}$ are strongly desired. 

We should also mention the relatively high $T_{\rm HP}$.
The present neutron data and the previous M\"ossbauer data\cite{Xu2} showed the temperature, where the magnetic SRO appears, increases drastically.
As mentioned above, the lattice is contracted by the pressure e.g. the lattice is contracted by 2 \% along the $a$ axis and 0.5 \% along $c$ axis at 7.66 GPa.
It is naturally considered that the overall exchange interactions are enhanced by the contraction.
We infer that the drastic increment of $T_{\rm HP}$ under the pressure might be caused by the overall increment of the exchange interactions.
On the other hand, Takahashi et al. reported the slightly decreasing $T_{\rm N1}$ and $T_{\rm N2}$ with increasing pressure, by their magnetic susceptibility measurements.\cite{Takahashi}
Since both the PD and 4SL states are stabilized with help of the lattice distortion, those states should be suppressed by the pressure as mentioned above.
The further studies for measurements under pressure varied systematically would be variable to test these predictions.

There have been several studies of the effect of pressure on magnetic ordering in spin-lattice coupling systems.
The chromium spinels are a well known frustrated spin system with spin-lattice coupling.\cite{Matsuda,Tanaka}
Ueda {\it et al.} reported the effect of pressure on magnetic phase transitions in the chromium spinels CdCr$_{2}$O$_{4}$, HgCr$_{2}$O$_{4}$, and ZnCr$_{2}$S$_{4}$.\cite{Ueda_P}
Although the direct exchange interaction was enhanced and the critical fields were changed by the pressure, no drastic change was observed in the spinel compounds.
For Tb$_{2}$Ti$_{2}$O$_{7}$ with a spin liquid ground state,\cite{Gardner} Mirebeau {\it et al.} reported the pressure-induced crystallization of a spin liquid.\cite{Mirebeau}
The present study of CuFeO$_{2}$ provides a rare example demonstrating the pressure-suppression of LRO in a frustrated spin lattice coupling system.

In conclusion, we studied the effect of pressure on magnetic ordering  in frustrated magnetic systems through neutron diffraction experiments under an isotropic pressure in the triangular lattice antiferromagnet CuFeO$_{2}$.
We found that the magnetic LRO in the four-sublattice ground state under ambient pressure in CuFeO$_{2}$ is suppressed at 7.9 GPa, and is replaced by an incommensurate SRO with a temperature-independent wave number of (0.192 0.192 1.5).
Considering the previous M\"ossbauer spectroscopy measurements under pressure, we concluded that the observed incommensurate SRO is a static order. 
The incommensurate wave number observed at 7.9 GPa corresponds to that just above the temperature at which the lattice distortion and magnetic LRO simultaneously occur under ambient pressure.
We therefore conclude that the long range magnetic ordering disappears when the high pressure suppressed the lattice distortion that relieves the spin frustration and leads the spin system to LRO.
Finally, since CuFeO$_{2}$ is a unique example of a spin-lattice coupling system in which the magnetic LRO is suppressed by pressure, the present study provided a good opportunity to study spin-lattice coupling in a geometrically frustrated magnetic system.

This work was partly supported by Grants-in-Aid for Scientific Research ``Young Scientists (B), Grants No. 2074209" from JSPS.


\end{document}